\documentclass[
twocolumn,
amsmath,
aps,
prb]{revtex4-1}
\usepackage[utf8]{inputenc}
\usepackage{graphicx}
\usepackage{dcolumn}
\usepackage{bm}
\usepackage{mathrsfs,ulem}
\usepackage[pdftex,
            colorlinks=true,
            citecolor=blue,
            linkcolor=blue]{hyperref}

\begin{document}
\title{Intrinsic superconducting diode effect in disordered systems}

\author{Yuhei Ikeda}
\author{Akito Daido}
\author{Youichi Yanase}
\affiliation{Department of Physics, Graduate School of Science, Kyoto University, Kyoto 606-8502, Japan}

\date{\today}

\begin{abstract}
Nonreciprocal transport phenomena have attracted much attention in modern condensed matter physics. In the field of superconductivity, the superconducting diode effect (SDE) has been one of the central topics. Recent theoretical studies for the SDE in intrinsic mechanism revealed the relation between the SDE and helical superconductivity, for which experimental clarification has been awaited. In this work, we establish a microscopic theory of the intrinsic SDE in disordered systems. We show that the sign reversal of the nonreciprocal critical current is suppressed under moderate impurity concentrations. However, even in the moderately disordered region, the SDE shows a feature signaling the change in the nature of helical superconductivity. It is also found that the diode quality factor $r$ is increased by disorders and reaches 20\% in the Rashba-Zeeman model.
\end{abstract}

\maketitle

\section{\label{sec:Intro}Introduction}
In recent years, nonreciprocal transport phenomena have attracted much attention in the field of condensed matter physics as a new functionality of noncentrosymmetric matter~\cite{Tokura2018-lp,Ideue2021-vu}.
Nonreciprocal transport means inequivalence between transport responses in one direction and the opposite direction, and phenomena such as magneto-chiral anisotropy have been widely known for a long time~\cite{Rikken1997-sa,Rikken2001-oo}.
In superconductors, the discovery and prediction of numerous nonreciprocal phenomena, including the early discovery of magneto-chiral anisotropy in the fluctuation region~\cite{Wakatsuki2017-bc,Wakatsuki2018-fs,Hoshino2018-yr,Qin2017-vd}, have been reported and are expanding~\cite{Yasuda2019-gb,Itahashi2020-tg,Zhang2020-al,Itahashi2022-rj,Wu2022-kw,Nakamura2020-vj}.
Recently, it has been shown that nonreciprocity also appears in the critical current in Nb/V/Ta artificial superlattice superconductors~\cite{Ando2020-ge}, and the nonreciprocal critical current is called the superconducting diode effect (SDE).
The discovery of the SDE in the Nb/V/Ta superlattice has led to an explosion of related research~\cite{Miyasaka2021-ie,Narita2022-ig,Kawarazaki2022-pd,Lin2022-md,Wu2022-vw,Xie2022-jo,Shin2021-gc,Scammell2022-mq,Baumgartner2022-gg,Bauriedl2022-mf,Lu2022-mv,Legg2022-gz,Chahid2022-pq,Jiang2022-ud,Hou2022-le,Pal2021-ry,Davydova2022-fn,Zinkl2021-dn,Lyu2021-ki,Hu2022-zv,Anwar2022-on}, and experimental observations have been reported not only in the original Rashba superconductors with an in-plane magnetic field but also in the Ising superconductors~\cite{Bauriedl2022-mf}, Josephson junction~\cite{Wu2022-vw,Baumgartner2022-gg}, and various systems even without the magnetic field~\cite{Wu2022-vw,Narita2022-ig,Lin2022-md}.
Diodes are fundamental elements of modern electronics, and therefore, the establishment of the SDE is expected to lead to energy-saving electronics using superconducting circuits.

Theories of the SDE have been formulated in terms of an intrinsic deparing mechanism \cite{Daido2022-wb,He2022-bc,Yuan2022-xh,Daido2022-gp,Ilic2022-la}.
Depairing of Cooper pairs by the supercurrent destabilizes the superconductivity and causes the depairing critical current.  
At the depairing critical current, the condensation energy due to the creation of Cooper pairs is compensated by the kinetic energy of Cooper pairs. It was shown that asymmetry in the band structure induces nonreciprocal properties in the depairing critical current in the sense that the critical current in a direction $J_\mathrm{c}(+)$ is different from that in the opposite direction $J_\mathrm{c}(-)$.
Theories have revealed the scaling of the nonreciprocal critical current $\Delta J_\mathrm{c} \equiv J_\mathrm{c}(+)-J_\mathrm{c}(-) \propto (T_\mathrm{c}-T)^2$. 
More interestingly, microscopic calculations predict the significant enhancement of the SDE beyond the $(T_\mathrm{c}-T)^2$ scaling in the low-temperature region and the sign reversal of SDE in the high-magnetic-field region~\cite{Daido2022-wb}.
The sign-reversal field roughly coincides with the crossover line of the helical superconducting state~\cite{Daido2022-wb,Daido2022-gp,Ilic2022-la}, indicating the close relationship between the SDE and finite-$q$ Cooper pairing. 
Therefore, developments in the intrinsic SDE are also desired for the clarification of exotic helical superconductivity, which has been awaited for a long time~\cite{Smidman2017-hv}. In particular, further theoretical research is desired for a comprehensive understanding of the SDE.

\begin{figure}[t]
    \centering
    \includegraphics[width = 0.48\textwidth]{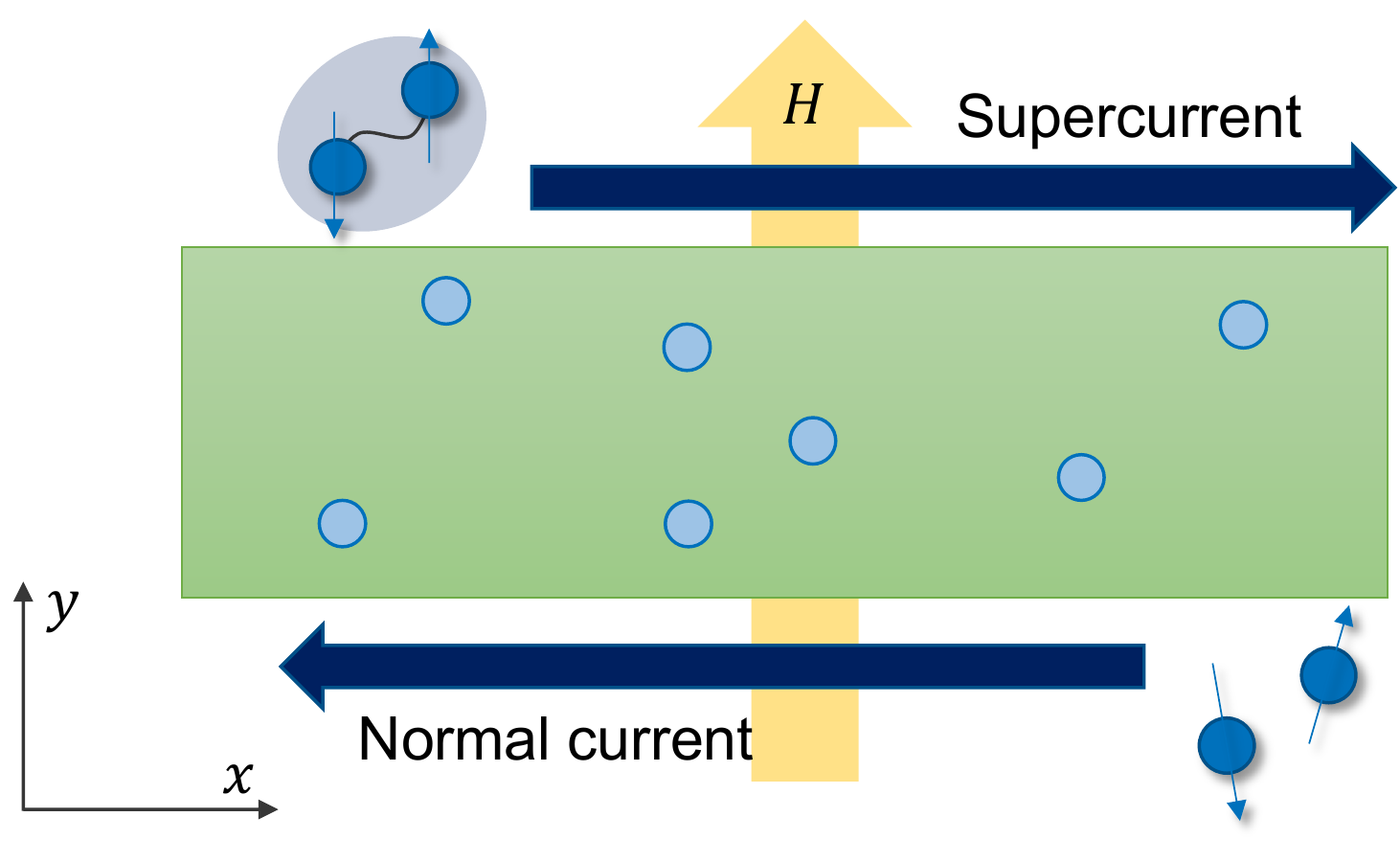}
    \caption{Schematic figure of the SDE with nonmagnetic impurities. Electric current flows in the $+x$ direction with zero resistance, while the electric resistance is finite in the $-x$ direction. The rightward current is the supercurrent and the leftward current is the normal current.}
    \label{fig:Schematic}
\end{figure}

Most of the theoretical calculations for the SDE were performed in the clean limit.
However, it is known that disorders have a significant effect on noncentrosymmetric superconductors~\cite{Smidman2017-hv,Dimitrova2007-gl,Samokhin2008-pk,Mineev2007-ut,Houzet2015-cl}.
In a previous study~\cite{Ilic2022-la}, impurity effects on the SDE near the transition temperature were studied using Ginzburg-Landau (GL) theory with the GL coefficients determined by the quasiclassical theory, which revealed that the sign of $\Delta J_\mathrm{c}$ depends on the impurity concentration. 
On the other hand, the intrinsic SDE in disordered systems has not been studied in the low-temperature region. The intrinsic SDE in the low-temperature region is attractive because it may realize high SDE efficiency~\cite{Daido2022-wb,Daido2022-gp} and experimental studies of helical superconductivity are feasible.
Therefore, we present a microscopic study of the SDE in disordered $s$-wave superconductors with a particular interest in the magnitude and sign reversal of the SDE in the low-temperature region.

In this paper, we investigate the nonreciprocal critical current with nonmagnetic impurities, based on a Green's function method.
In particular, we focus on the weakly-disordered regime since the crossover of helical superconductivity is known to be absent in the presence of strong disorder~\cite{Smidman2017-hv}.
In Sec.~\ref{Model}, we show a formulation of the SDE in the disordered Rashba-Zeeman model for $s$-wave superconductors. 
In Sec.~\ref{sec:Result}, we show the numerical results of the SDE. It is shown that the sign reversal of SDE observed in the clean limit is suppressed with increasing impurity concentration.
It is also found that the diode quality factor is increased by disorders and reaches 20\%. 

\section{Formulation}\label{Model}

In this section, we describe a microscopic treatment of SDE.
We first derive Green's function method for clean systems and next extend it to disordered systems. 
A setup of the system is shown schematically in Fig.~\ref{fig:Schematic}.
We consider two-dimensional superconductors with a polar axis perpendicular to the conducting plane. 
Following the setup of Ref.~\onlinecite{Daido2022-wb}, we adopt the Rashba-Zeeman Hamiltonian with an attractive Hubbard interaction,
\begin{align}
    \mathcal{H} &= \mathcal{H}_0 + \mathcal{H}_\mathrm{int}, \\
    \mathcal{H}_0 &= \sum_{\bm{k}\sigma\sigma'}[(\varepsilon(\bm{k})-\mu)\delta_{\sigma\sigma'}+(\bm{g}(\bm{k})- \bm{h})\cdot \bm{\sigma}_{\sigma\sigma'}]c^\dag_{\bm{k}\sigma}c_{\bm{k}\sigma'},  \\
    \mathcal{H_\mathrm{int}} &=-\frac{U}{V}\sum_{\bm{k k'q}}c^\dag_{\bm{k_+}\uparrow}c^\dag_{-\bm{k_-}\downarrow}c_{-\bm{k'_-}\downarrow}c_{\bm{k'_+}\uparrow}. 
\end{align}
Here, $\varepsilon(\bm{k}) = -2t(\cos k_x+\cos k_y)$ is a nearest-neighbor hopping energy in the tight-binding model, $\mu$ is a chemical potential, and $\bm{g}(\bm{k}) = \alpha_\mathrm{R}(-\sin k_y, \sin k_x,0)$ represents the Rashba spin-orbit coupling. 
The magnetic field is applied along the $y$-axis, treated as a Zeeman term $\bm{h} = (0,\mu_{\mathrm{B}}H,0)$.
We introduce the notation $\bm{k}_{\pm} = \bm{k} \pm \bm{q}/2$, and the system size is $V = L_x L_y$. The parameters are given by $( t, \mu, \alpha_\mathrm{R}, U) = (1.0,-1.0,0.3,1.5)$. 

The energy dispersion for the non-interacting part is $\xi_\lambda(\bm{k},\bm{h}) = \varepsilon(\bm{k})-\mu +\lambda | \bm{g}(\bm{k}) -\bm{h}|$, where $\lambda$ represents the helicity $\lambda = \pm$. 
Unitary transformation to the helicity band is described as follows,
\begin{equation}
    c_{\bm{k}\sigma} = \sum_{\lambda} u_{\sigma\lambda}(\bm{k},\bm{h})c_{\bm{k}\lambda},
\end{equation}
with the unitary matrix $\left(u_{\uparrow\lambda}(\bm{k},\bm{h}),u_{\downarrow\lambda}(\bm{k},\bm{h})\right) = 1/\sqrt{2} (1,\lambda e^{i\phi(\bm{k},\bm{h})})$ and $\phi(\bm{k},\bm{h}) =$ 
{$\arg[g_x(\bm{k})-h_x+i(g_y(\bm{k})-h_y)]$.}
For convenience of notation, we define $t_\lambda(\bm{k},\bm{h}) \equiv \lambda e^{i\phi(\bm{k},\bm{h})}$.
We apply the same unitary transformation to the interaction part,
\begin{align}
    -\frac{U}{4V} \sum_{\substack{\bm{k,k'}\\ \lambda_1\lambda_2\lambda_3\lambda_4}} t_{\lambda_2}(-\bm{k}_-,\bm{h})t^*_{\lambda_3}(-\bm{k}'_-,\bm{h}) \notag\\
    \times c^\dag_{\bm{k}\lambda_1}c^\dag_{-\bm{k}_-\lambda_2}c_{-\bm{k'}_-\lambda_3}c_{\bm{k'}\lambda_4}.
\end{align}

We solve this model in the mean-field approximation with finite center-of-mass Cooper-pair momentum $\bm{q}$. In this setup nondissipative supercurrent is carried by Cooper pairs with finite momentum. Then, the Hubbard interaction term is decoupled by the gap function defined as follows,
\begin{equation}
    \tilde{\Delta} (q) = -\frac{U}{2V}\sum_{\bm{k}'\lambda'} t^*_{\lambda'}(-\bm{k}'_-,\bm{h})\langle c_{-\bm{k'}_-\lambda'}c_{\bm{k'}_+\lambda'}\rangle,
\end{equation}
by neglecting the interband pairing for simplicity.
We consider $s$-wave paring with center-of-mass momentum along the $x$ direction $\bm{q} =q\hat{\bm{x}}$. 
Using the above treatment, we obtain the interaction part in the mean-field Hamiltonian
\begin{equation} \mathcal{H}_{\mathrm{int}} \simeq \frac{1}{2}\sum_{\bm{k}\lambda} \left(\Delta_\lambda (\bm{k},\bm{h},q) c^\dag_{\bm{k}_+\lambda}c^\dag_{-\bm{k}_-\lambda} + \mathrm{H.c.} \right) + \mathrm{Const.}
\end{equation}
Here, we denote $\Delta_\lambda(\bm{k},\bm{h},q) =  t_\lambda (-\bm{k}_-,\bm{h})\tilde{\Delta}(q)$, 
which can be replaced with its antisymmetric part $\Delta_{0\lambda}(\bm{k},\bm{h},q)=\left(\Delta_{\lambda}(\bm{k},\bm{h},q)-\Delta_{\lambda}(-\bm{k},\bm{h},q)\right)/2$ according to the anticommutation relation of creation operators.
Thus, in the band basis, the mean-field Hamiltonian is expressed as follows,
\begin{align}
    \mathcal{H}_{\mathrm{MF}} &= \sum_{\bm{k}\lambda}\xi_{\lambda}(\bm{k},\bm{h})c^\dag_{\bm{k}\lambda}c_{\bm{k}\lambda} \notag \\
    &+ \frac{1}{2}\sum_{\bm{k}\lambda} \left(\Delta_{{0}\lambda} (\bm{k},\bm{h},q) c^\dag_{\bm{k}_+\lambda}c^\dag_{-\bm{k}_-\lambda} + \mathrm{H.c.} \right) + \mathrm{Const.}
\end{align}

The explicit notation of the magnetic-field dependence is omitted below. We now introduce Green's function on the particle-hole basis as follows;
\begin{equation}
    \mathscr{G}(\bm{k},\tau) = -\langle T_\tau C_{\bm{k}}(\tau)C^{\dag}_{\bm{k}}(0) \rangle.
\end{equation}
Where, $C_{\bm{k}} = (c_{\bm{k}_++},c_{\bm{k}_+-},c^{\dag}_{-\bm{k}_-+},c^{\dag}_{-\bm{k}_--})^T$.
In each component, the Fourier transform is defined using fermion's Matsubara frequency $\omega_n$ as follows;
\begin{equation}
    {\mathscr{G}}(\bm{k},\tau) = \frac{1}{\beta}\sum_{\omega_n} {\mathscr{G}}(\bm{k},\omega_n) e^{-i\omega_n \tau}.
\end{equation}
From this, Green's function in the particle-hole basis is represented by
\begin{equation}
    \mathscr{G}(\bm{k},\omega_n) = \begin{bmatrix}
      G({\bm{k}_+},\omega_n)  &   -F(\bm{k},\omega_n) \\
      -F^\dag(\bm{k},\omega_n)  &   -G^T(-{\bm{k}_-},-\omega_n)
    \end{bmatrix}.
\end{equation}
In the equation, the normal and anomalous Green's functions are defined as follows;
\begin{align}
    G_{\lambda\lambda'}(\bm{k},\tau) &= -\langle T_{\tau} c_{{\bm{k}}\lambda}(\tau)c^\dag_{{\bm{k}}\lambda'}(0) \rangle, \\
    F_{\lambda\lambda'}(\bm{k},\tau) &=\langle T_{\tau} c_{\bm{k}_+\lambda}(\tau)c_{-\bm{k}_-\lambda'}(0) \rangle.
\end{align}
{These Green's functions are proportional to  $\delta_{\lambda\lambda'}$ by the assumption of the vanishing interband pairing.
Thus, we write $G_{\lambda\lambda'}(\bm{k},\tau)=G_\lambda(\bm{k},\tau)\delta_{\lambda\lambda'}$ and so on.}
After some calculations, we obtain Gor'kov equations,
\begin{align}
    (i\omega_n - \xi_{\lambda}(\bm{k}_+)) G_\lambda({\bm{k}_+},\omega_n) +\Delta_{0\lambda} F^\dag_\lambda(\bm{k},\omega_n) &= 1, \\
    (i\omega_n + \xi_\lambda(-\bm{k}_-))F^\dag_\lambda(\bm{k},\omega_n) + \Delta^\dag_{0\lambda}G_\lambda({\bm{k}_+},\omega_n)&=0.
\end{align}
%
The normal and anomalous Green's functions are obtained as 
\begin{align}
    G_{\lambda}&({\pm\bm{k}_\pm},\omega_n) \notag\\
    &= \frac{i\omega_n + \xi_\lambda(\mp\bm{k}_\mp)}{(i\omega_n - \xi_\lambda(\pm\bm{k}_\pm))(i\omega_n + \xi_\lambda(\mp\bm{k}_\mp)) - |\Delta_{0\lambda}|^2},
\end{align}
and
\begin{align}
    F_{\lambda}&(\bm{k},\omega_n) \notag \\
    &= -\frac{\Delta_{0\lambda}(\bm{k},q)}{(i\omega_n - \xi_\lambda(\bm{k}_+))(i\omega_n + \xi_\lambda(-\bm{k}_-)) - |\Delta_{0\lambda}|^2}.
\end{align}

Then, we take into account the impurity effect. Here, we consider scattering by nonmagnetic impurities,
\begin{equation} \mathcal{H}_{\mathrm{imp}} = \frac{1}{V}\sum_{\bm{k}\bm{k'}\sigma} U_{\mathrm{imp}} (\bm{k} - \bm{k}') c^\dag_{\bm{k}\sigma}c_{\bm{k'}\sigma}.
\end{equation}
This impurity Hamiltonian is described in the band basis as follows,
\begin{align}
& \mathcal{H}_{\mathrm{imp}} = \frac{1}{V}\sum_{\substack{\bm{k}\bm{k'}\\ \lambda\lambda'}} U_{\mathrm{imp}} (\bm{k} - \bm{k}') w_{\lambda \lambda'}(\bm{k},\bm{k'})c^\dag_{\bm{k}\lambda}c_{\bm{k'}\lambda'}, \\
& w_{\lambda \lambda'}(\bm{k},\bm{k'}) = \sum_{\lambda''}u_{\lambda\lambda''}^*(\bm{k})u_{\lambda''\lambda'}(\bm{k}').
\end{align}
We assume a random distribution of isotropic impurities with $U_{\mathrm{imp}} (\bm{k} - \bm{k}') = U_0$ and a concentration $n_{\mathrm{imp}}$ and coarse-grain impurities. By applying the self-consistent Born approximation, we obtain the following equation,
\begin{equation}
    (\check{\mathcal{G}}_0^{-1} - \check{\Sigma}_{\mathrm{imp}})\check{\mathcal{G}} = \check{1}.
\end{equation}
By solving this equation, we obtain the Green's function and impurity self-energy by,
\begin{align}
\check{\mathcal{G}}^{-1}_0 (\bm{k},\omega_n) &= \begin{bmatrix}
    i\omega_n - \hat{\xi}(\bm{k}_+) & -\hat{\Delta}_0\\
    -\hat{\Delta}_0^\dag & i\omega_n + \hat{\xi}(-\bm{k}_-) 
    \end{bmatrix}, \\
    \check{\Sigma}_{\mathrm{imp}}(\bm{k},\omega_n) &= \frac{n_{\mathrm{imp}}U_0^2}{V} \sum_{\bm{k}'}\check{W}(\bm{k},\bm{k}')\check{\mathcal{G}}(\bm{k}',\omega_n)\check{W}(\bm{k}',\bm{k}),
\end{align}
where
\begin{align}
    \check{W}(\bm{k},\bm{k'}) &= \begin{bmatrix}
      \hat{w}(\bm{k}_+,\bm{k'}_+) & 0 \\
      0 & -\hat{w}^T(-\bm{k}'_-,-\bm{k}_-)
    \end{bmatrix}.
\end{align}
In the absence of impurities, Green's function is band-diagonal. 
Therefore, we take account of the intraband scattering. 
Based on this requirement, we calculate the impurity self-energy as
\begin{align}
\check{\Sigma}_{\mathrm{imp}}(\omega_n) = \begin{bmatrix}
    \hat{\Sigma}^+_1(\omega_n) &\hat{\Sigma}_2(\omega_n) \\
    \hat{\Sigma}_2^*(\omega_n) & -\hat{\Sigma}^-_1(-\omega_n)
    \end{bmatrix},
\end{align}
where each component is represented by
\begin{align}
    \hat{\Sigma}_1^\pm(\omega_n) &= \frac{n_{\mathrm{imp}}U_0^2}{2V}\hat{\tau}_0\sum_{\bm{k}\lambda} {G_{\lambda}(\pm\bm{k}_\pm,\omega_n),}\\
    \hat{\Sigma}_2(\omega_n) &= \frac{n_{\mathrm{imp}}U_0^2}{2V}\hat{\tau}_0\sum_{\bm{k}\lambda} F_{\lambda}(\bm{k},\omega_n).
\end{align}
Here, $\hat{\tau}_0$ represents the unit matrix in the helicity band basis.  By solving the Gor'kov equation, the impurity averaged Green's functions are expressed as
\begin{align}
    G_{\lambda}&({\pm}\bm{k}_\pm,\omega_n) = \notag\\
    & \frac{i\tilde{\omega}_n + \xi_\lambda({\mp}\bm{k}_\mp)}{(i\tilde{\omega}_n - \xi_\lambda({\pm}\bm{k}_{{\pm}}))(i\tilde{\omega}_n + \xi_\lambda({\mp}\bm{k}_{{\mp}})) - |D_\lambda(\omega_n)|^2},
\end{align}
\begin{align}
    F_{\lambda}&(\bm{k},\omega_n) = \notag \\
    & -\frac{D_\lambda(\omega_n)}{(i\tilde{\omega}_n - \xi_\lambda(\bm{k}_+))(i\tilde{\omega}_n + \xi_\lambda({-}\bm{k}_-)) - |D_\lambda(\omega_n)|^2}.
\end{align}
We include the real part of $\Sigma_1$ into the chemical potential and take the gap function to be real without loss of generality. The impurity effects are obtained by solving the following self-consistent equations,
\begin{widetext}
\begin{align}\label{eq:self-born1}
    \tilde{\omega}_n &= \omega_n + \frac{n_{\mathrm{imp}}U^2_0}{2V} \sum_{\bm{k}\lambda}\frac{\tilde{\omega}_n(\tilde{\omega}^2_n+|D_\lambda(\omega_n)|^2+\xi^2_\lambda(-\bm{k}_-))}{(\tilde{\omega}_n^2+|D_\lambda(\omega_n)|^2+\xi_\lambda(\bm{k}_+)\xi_\lambda(-\bm{k}_-))^2+\tilde{\omega}_n^2(\xi_\lambda(\bm{k}_+)-\xi_\lambda({-}\bm{k}_-))^2}, \\
    D_\lambda(\omega_n) &= \Delta_{0\lambda}(q) +\frac{n_{\mathrm{imp}}U^2_0}{2V}\sum_{\bm{k}\lambda}\frac{D_\lambda(\omega_n)(\tilde{\omega}^2_n+|D_\lambda(\omega_n)|^2+\xi_\lambda(\bm{k}_+)\xi_\lambda(-\bm{k}_-))}{(\tilde{\omega}_n^2+|D_\lambda(\omega_n)|^2+\xi_\lambda(\bm{k}_+)\xi_\lambda(-\bm{k}_-))^2+\tilde{\omega}_n^2(\xi_\lambda(\bm{k}_+)-\xi_\lambda(-\bm{k}_-))^2}.\label{eq:self-born2}
\end{align}
\end{widetext}
Solving the self-consistent equations numerically, we obtain the normal and anomalous Green's functions, impurity self-energy, and gap function as functions of the Cooper-pair momentum $q$.
The electric current is calculated by using the Green's function as follows,
\begin{align}
    J_x(q) &= \frac{1}{V}\sum_{\bm{k}n\lambda}v_{\lambda}({\bm{k}_+})G_\lambda({\bm{k}_+},\omega_n), \\
    v_\lambda(\bm{k})&=\partial_{k_x}\xi_\lambda(\bm{k}).
\end{align}
The obtained electric current is the supercurrent without dissipation because the electric field is zero in the above formulation. 
Therefore, the superconducting critical current corresponds to the maximum (minimum) value of the electric current and is calculated by
\begin{align}
    J_\mathrm{c}(+) &= \max_{q}J_x(q), \\
    J_\mathrm{c}(-) &= \min_{q}J_x(q).
\end{align}
Consequently, the nonreciprocity in the superconducting critical current is defined as
\begin{equation}\label{eqs:diode}
    \Delta J_\mathrm{c} \equiv J_\mathrm{c}(+) - |J_\mathrm{c}(-)|.
\end{equation}
Using these equations, the diode quality factor $r$, an index for evaluating the performance of the SDE, can be expressed as follows,
\begin{equation}\label{eqs:r}
r \equiv \frac{\Delta J_\mathrm{c}}{2\bar{J_\mathrm{c}}} = \frac{\Delta J_\mathrm{c}}{J_\mathrm{c}(+) {+} |J_\mathrm{c}(-)|}.
\end{equation}
Since no current flows in equilibrium, the Cooper-pair momentum $q_0$ in the stable helical superconducting state is obtained by the equation
\begin{equation}\label{eq:helical}
    J_x(q_0) = 0.
\end{equation}

\section{Nonreciprocal critical current}\label{sec:Result}
In this paper, we focus on the SDE in the low-temperature region~\cite{Daido2022-wb,Daido2022-gp}.
The main features of the intrinsic SDE in the clean limit are the enhanced SDE just below and within the crossover regime of helical superconductivity as well as the sign reversal(s) in the crossover region. We are interested in whether and how these properties are modified in moderately-clean and weakly-disordered noncentrosymmetric superconductors.
Note that the large SDE just below the onset of the crossover is a unique feature of the low-temperature SDE and has not been explored in disordered systems so far.
Such behavior is absent or subtle near the transition temperature due to the vanishing $H$-linear SDE in the GL regime of $s$-wave superconductors~\cite{Ilic2022-la}.

\begin{figure}[t]
    \centering
\includegraphics[width=0.48\textwidth]{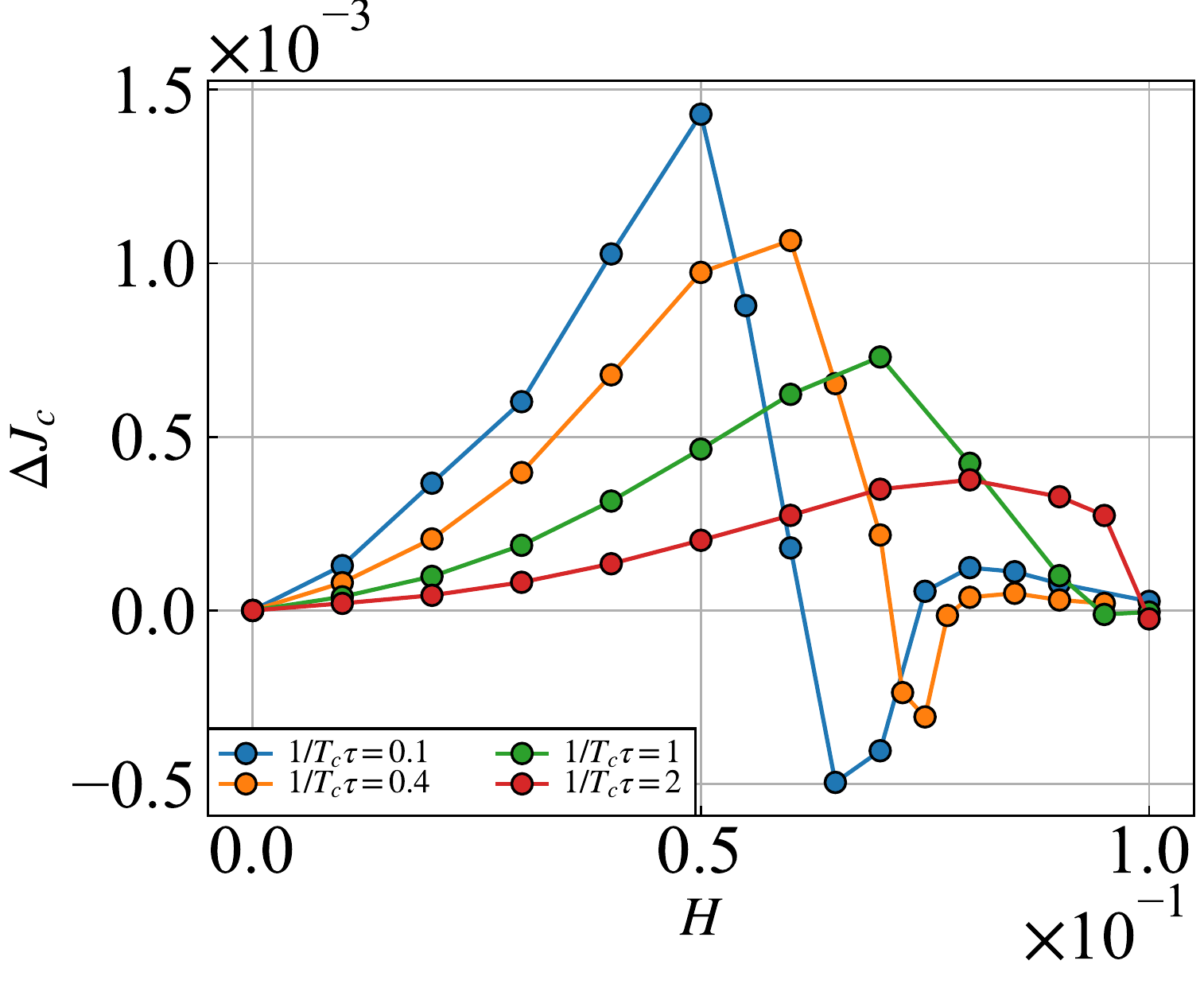}
    \caption{The magnetic field dependence of the nonreciprocal critical current $\Delta J_\mathrm{c}$ at $T=0.01$, which is much lower than the transition temperature $T_\mathrm{c} \simeq 0.0389$. Each colored line with circles represents $\Delta J_\mathrm{c}(H)~$ for impurity concentrations corresponding to $(T_\mathrm{c} \tau)^{-1}=0.1$, $0.4$, $1$, $2$.
    }
    \label{fig:H_diode}
\end{figure}

To study the low-temperature SDE in disordered systems, we set $T = 0.01$ in this paper, which is much lower than the transition temperature $T_\mathrm{c} \simeq 0.0389$ in our unit $t=1$. 
We show in Fig.~\ref{fig:H_diode} the magnetic-field dependence of the nonreciprocal critical current $\Delta J_\mathrm{c}$ [Eq.~\eqref{eqs:diode}] for several impurity concentrations. 
The impurity effects are parameterized by $(\tau T_\mathrm{c})^{-1}$ as usual, where $\tau$ is the lifetime in the normal state defined by
\begin{align}
\tau^{-1} = \pi n_{\mathrm{imp}}U^2_0N_\mathrm{F},
\end{align}
where $N_{\mathrm{F}}=\frac{1}{2V}\sum_{\bm{k}\lambda}\delta[\xi_{\lambda}(\bm{k})]$ is the density of states at the Fermi level.

We first discuss the moderately clean case, 
$(\tau T_\mathrm{c})^{-1}=0.1$ in Fig.~\ref{fig:H_diode}. 
The nonreciprocal critical current increases with the magnetic field for $H<0.05$, and it shows a peak, reaching a maximum value at a magnetic field $H=0.05$. 
When the magnetic field further increases, the nonreciprocal critical current  rapidly decreases, and the sign turns negative. As approaching the upper critical field, the magnitude of the nonreciprocal critical current gradually decreases. 
This magnetic-field dependence of the SDE is in good agreement with the previous studies in the clean limit~\cite{Daido2022-wb,Daido2022-gp,Ilic2022-la}.
In particular, the sign change in the SDE is robust against a small impurity concentration.
Therefore, the sign reversal phenomenon in the intrinsic SDE should be observed not only in the clean limit but also in moderately-clean superconductors.

When the impurity concentration is further increased, both $|J_\mathrm{c}(+)|$ and $|J_\mathrm{c}(-)|$ generally become small, and therefore $\Delta J_\mathrm{c}$ also gets suppressed as an overall tendency [$1/T_\mathrm{c}\tau=0.4,\,1$ and $2$ in Fig.~\ref{fig:H_diode}].
Nevertheless, the SDE is still enhanced by increasing the magnetic field as in the moderately-clean case. Thus, the peaked SDE under the moderate magnetic field is clarified to be a general feature of clean and weakly-disordered Rashba superconductors.
Another intriguing feature in the weakly-disordered region is the sign-reversing region of $\Delta J_\mathrm{c}(H)$ becomes hardly visible: After reaching its peak, $J_\mathrm{c}(H)$ is suppressed and never takes a large value with the opposite sign in contrast to the moderately-clean case.

\begin{figure}[t]
    \centering
\includegraphics[width=0.48\textwidth]{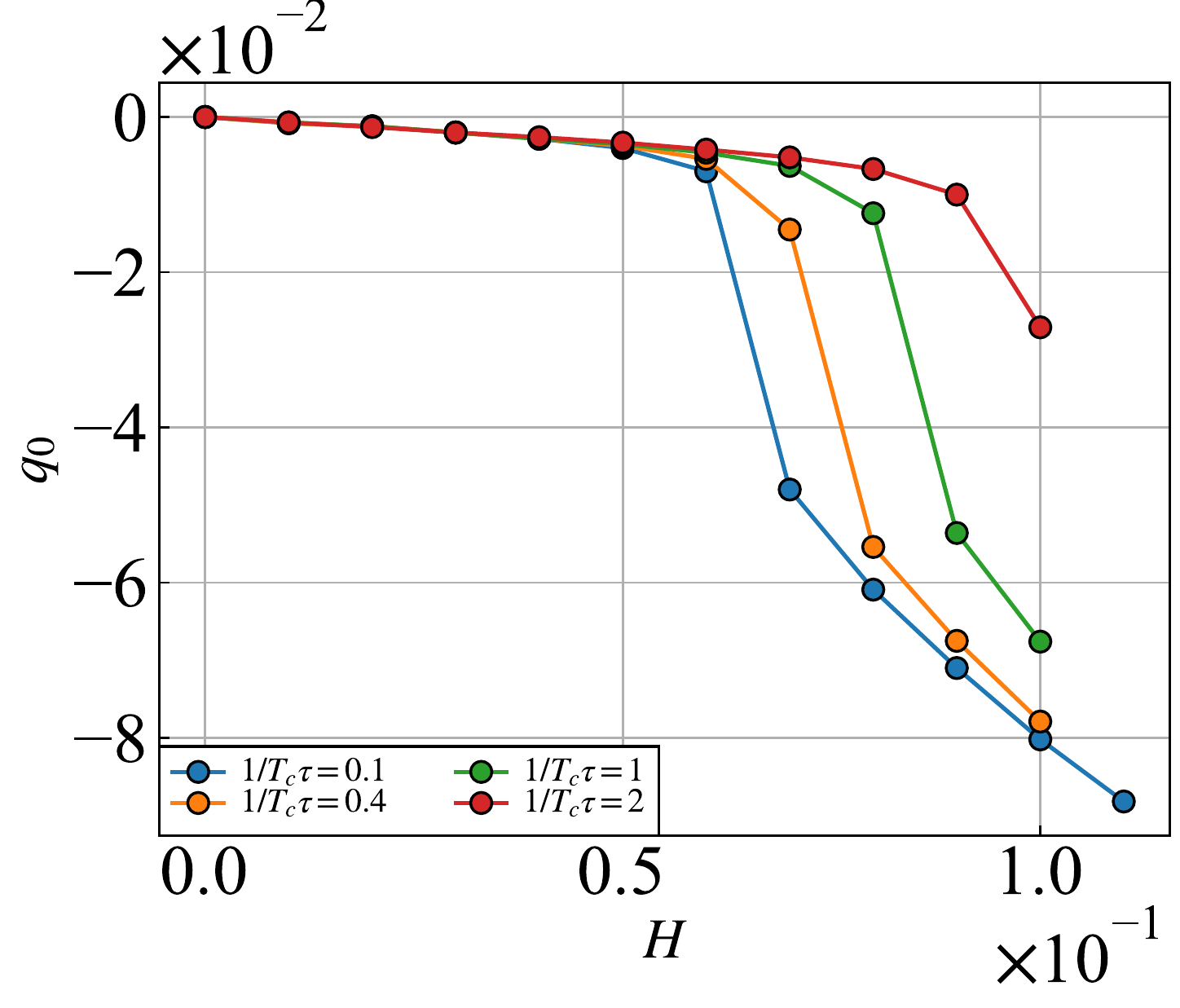}
    \caption{The magnetic field dependence of the Cooper-pair momentum $q_0$ in the equilibrium state at $T=0.01$.
    }
    \label{fig:helical}
\end{figure}

In order to consider the relationship between the SDE and helical superconductivity pointed out in Ref.~\onlinecite{Daido2022-wb}, we show in Fig.~\ref{fig:helical} the magnetic-field dependence of the Cooper-pair momentum $q_0$ in equilibrium [Eq.~\eqref{eq:helical}]. 
For a small impurity concentration $(T_\mathrm{c} \tau)^{-1}=0.1$, the Cooper-pair momentum $q_0$ shows a rapid increase above $H=0.06$. 
This is the crossover from the weakly to strongly helical states,
where the nature of helical superconductivity changes as increasing the magnetic field~\cite{Daido2022-wb,Smidman2017-hv,Dimitrova2007-gl,Kaur2005-aj}. The low-field phase is essentially two-band superconductivity, while the high-field phase is sustained by one of the Rashba-split Fermi surfaces.
Figure~\ref{fig:helical} shows the shift of the crossover field to the high-field region by increasing disorders. 
From the comparison of Figs.~\ref{fig:H_diode} and \ref{fig:helical}, we find that the starting point of the crossover coincides with the peak position of $\Delta J_\mathrm{c}$ shown in Fig.~\ref{fig:H_diode}. 
These results support the claim of the previous study that shows the close relationship 
between the giant intrinsic SDE and the crossover phenomenon~\cite{Daido2022-gp,Daido2022-wb,Ilic2022-la}.

\begin{figure}[t]
    \centering
\includegraphics[width=0.48\textwidth]{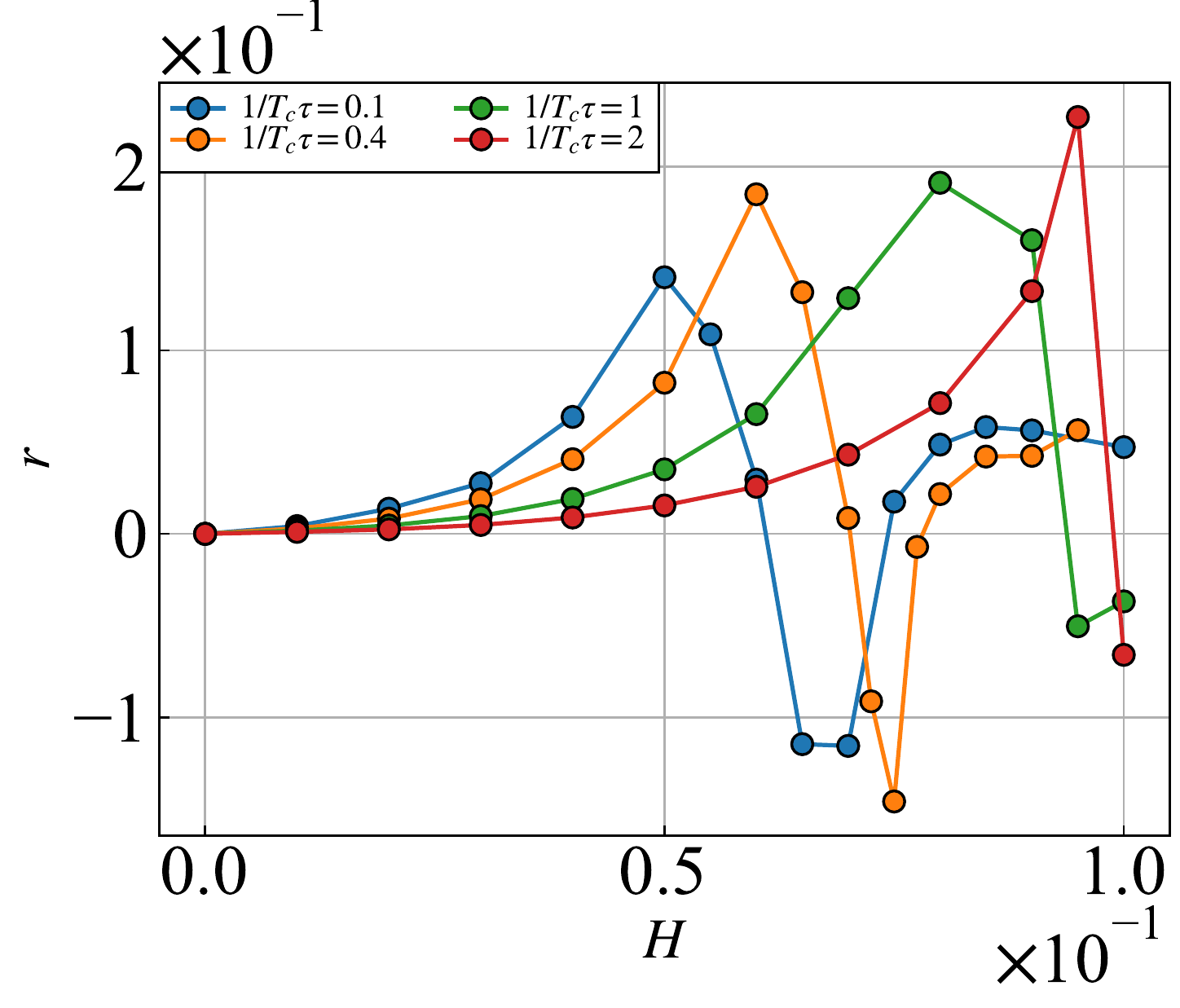}
    \caption{The magnetic field dependence of the SDE quality factor $r$ at $T=0.01$.
    }
    \label{fig:q-factor}
\end{figure}

Here we discuss the diode quality factor $r$ defined in Eq.~\eqref{eqs:r} (see Fig.~\ref{fig:q-factor}). 
As is clear from the definition, the diode quality factor shows a similar magnetic field dependence to the nonreciprocal critical current $\Delta J_\mathrm{c}$.
In particular, the peak position shifts to the higher magnetic field as the impurity concentration increases. 
Interestingly, the maximum value of $r$ increases with disorders in contrast to the case of $\Delta J_\mathrm{c}$.  This is because the averaged critical current is decreased by disorders more significantly than the nonreciprocal critical current. In this sense, the diode quality factor is enhanced by disorders, and it exceeds $r \simeq 20\%$ at $(\tau T_\mathrm{c})^{-1}=2$.
The stronger suppression of the averaged critical current $\bar{J}_c$ leads to a sizable diode quality factor even under high magnetic fields for weakly-disordered cases, in contrast to $\Delta J_\mathrm{c}$.
Thus, the sign reversals of the quality factor are observed for $1/T_\mathrm{c}\tau=1$ and $2$. 



\section{Summary and Discussion}
In this paper, we have studied the superconducting diode effect in $s$-wave noncentrosymmetric superconductors with nonmagnetic impurities.
We have derived the Gor'kov equation for the current flowing state in the Rashba-Zeeman model and taken into account the impurity effect by using the self-consistent Born approximation. 
We have revealed that the sign reversal of the diode effect is obtained not only in the clean limit but also in the moderately-clean systems, highlighting its feasibility as the probe of clean helical superconductors.
By further increasing the disorder strength,
on the other hand, the sign-reversed region gets suppressed. 
Thus, the characteristic magnetic field for the crossover phenomenon to occur can be detected by the intrinsic SDE even in the absence of the sign reversal.
We also found that the diode quality factor is enhanced by disorders. 
Since the diode quality factor is an index for the performance of SDE, our results indicate that a moderate impurity concentration optimizes the SDE.

Here we make a remark on the system studied in this paper. 
we calculated the SDE for impurity concentration from a nearly clean region $(\tau T_\mathrm{c})^{-1}=0.1$ to a moderately dirty region $(\tau T_\mathrm{c})^{-1}=2$ focusing on the disorder effects on the crossover of helical superconductivity.
The study of SDE in the dirty limit, where such crossover phenomenon is absent, is beyond the scope of this paper and is left for an intriguing future issue. 

In the end, we comment on the detection of helical superconductivity by using the SDE. 
The key factor in the detection of helical superconductivity is how to measure the spatially modulated order parameter as an observable.
One possible solution is to detect phase-sensitive phenomena such as the Josephson effect~\cite{Kaur2005-aj}.
Another possible detection may be observing phenomena related to the finite momentum of Cooper pairs, by controlling it using the supercurrent. 
Examples of the latter are the SDE and piezoelectric effect, which indeed drastically change by acquiring large Cooper-pair momentum~\cite{Daido2022-gp,Daido2022-wb,Chazono2022-rp}. 
When the intrinsic SDE is realized in clean noncentrosymmetric superconductors, the sign reversal of $\Delta J_\mathrm{c}$ is strong evidence of the helical crossover, namely, drastic change in the nature of helical superconductivity. 
The results presented in this paper indicate 
that the sign reversal might be difficult to observe in dirty superconductors compared with clean superconductors.
However, the peak in the nonreciprocal critical current signifies the helical-superconductivity crossover, and therefore, we can probe the helical superconductivity even in the moderately dirty region.  

\begin{acknowledgements}
We thank fruitful discussions with Teruo Ono, Yuta Miyasaka, Ryo Kawarazaki, Hideki Narita, Jun Ishizuka, and Shuntaro Sumita. YI thanks Yoshihiro Michishita for the technical advice. This work was supported by JSPS KAKENHI (Grants No.~JP18H01178, No.~JP18H05227, No.~JP20H05159, No.~JP21K18145, No.~JP21K13880, No.~JP22H04476, No.~JP22H01181, No.~JP22H04933), SPIRITS 2020 of Kyoto University, JSPS research fellowship, and WISE program MEXT.
\end{acknowledgements}
\bibliography{main}
\end{document}